\documentclass{article}
\usepackage{amsfonts,amssymb, amsmath}
\usepackage[english]{babel}
\usepackage{graphicx}
\usepackage{hyperref}
\usepackage{xcolor}

\newtheorem{thr}{Theorem}
\newtheorem{lem}{Lemma}

\begin{document}

\title{On the Chaplygin sphere in a magnetic field}
\author{A.V.Borisov$^{(1)}$, A.V. Tsiganov $^{(2)}$\\
$^{(1)}$ Udmurt State University;\\
$^{(2)}$ Steklov Mathematical Institute, Russian Academy of Sciences;\\
borisov@rcd.ru, andrey.tsiganov@gmail.com
}%
\date{}
\maketitle

\begin{abstract}
We consider the possibility of using Dirac's ideas of the deformation of Poisson brackets in nonholonomic mechanics.
As an example, we analyze the composition of external forces that do no work and reaction forces of nonintegrable constraints in the model of a nonholonomic Chaplygin sphere on a plane. We prove
that, when a solenoidal field is applied, the general mechanical energy, the invariant measure and the conformally Hamiltonian representation of the equations of motion are preserved. In addition, we consider the case of motion of the nonholonomic Chaplygin sphere in a constant magnetic field taking dielectric and ferromagnetic (superconducting) properties of the sphere into account. As a  by-product we also obtain two new integrable cases of the Hamiltonian rigid body dynamics in a constant magnetic field taking the magnetization by rotation effect into account.
\end{abstract}

\section{Introduction}
\setcounter{equation}{0}
\label{bor-sec1}

It has been known since as far back as the 1940s  that the equations of motion of a dielectric rigid body about a fixed point
in a homogeneous and stationary magnetic field are represented as Euler\,--\,Poisson equations with some generalized  potential.
This problem received particular attention in the work of Grioli~\cite{grioli47,grioli57,bb74} in which it was emphasized that
the work of external forces in this case is zero. For this reason it is often called the Grioli problem.

The equations of motion that appear in the Grioli problem are Hamiltonian
equations,  and one can also add to the corresponding Hamiltonian quadratic terms due to the Newtonian potential, the Brun field
or fast vibrations of the suspension point~\cite{bm01}. In this case, the equations have the form of Kirchhoff equations
describing the motion of a singly connected body in an infinite volume of an ideal incompressible fluid which is at rest at
infinity.
A detailed discussion of the integrability by quadratures of the Kirchhoff equations with various potential and gyroscopic terms
can be found in the book~\cite{bm01}.

If we add gyrostatic terms to the Grioli problem, then, from the viewpoint of motion in a fluid, they can be
treated as multiple connection of a rigid body, which admits circulation around contours. This analogy and the generalization of these
Hamiltonian equations to the $n$-dimensional case is discussed in detail in the book~\cite{bog91}.

Another, not less important, problem is that of the motion of a ferromagnetic or superconducting body in a magnetic field
with magnetization during rotation~\cite{ barn,beck, hir19,kob81,koz85,sam84,urm98}. In contrast to the Grioli problem,
in the general case the equations of motion are not Hamiltonian and, generally speaking, possess no invariant measure and
do not preserve the total mechanical energy of the body. After the change of variables, all well-known integrable particular cases~\cite{sam84,koz85}
in this problem are reduced either to the integrable Kirchhoff case~\cite{koz85} or to the integrable Clebsch case~\cite{ves86}
in the Kirchhoff equations, see Appendix D in the book~\cite{bm01}.

In this paper we consider the nonholonomic problem of the motion of an inhomogeneous dynamically balanced sphere rolling without
slipping on a horizontal plane with its center of mass at the geometric center. This is the so-called Chaplygin sphere.
Chaplygin found an invariant measure and first integrals for the equations of motion of the sphere in the gravitational field
and reduced the problem to quadratures. By the way, this year marks the 150th birthday of this great Russian mechanical engineer.
The generalization presented here is related to a sphere that exhibits both dielectric and ferromagnetic properties and
is acted upon by potential forces. We note that in the case of fast vibrations of the plane
a quadratic potential can appear and the study of such problems holds much promise for modern dynamics from the viewpoint of
vibrostabilization and vibrodisplacement.

 As a rule, the equations of motion of such systems cannot be represented in Hamiltonian
form, but can be reduced to it after conformal transformation of coordinates and time.
In addition to investigating the first integrals and invariant measures, we
present a general construction which makes it possible to include magnetic terms
in the conformally Hamiltonian representation of the equations of motion. Thus,
we present general cases of existence of additional integrals,
an invariant measure and a conformally Hamiltonian representation for
equations which, due to hydrodynamical analogy, can be called nonholonomic
Kirchhoff equations.

Other types of nonholonomic Kirchhoff equations with two degrees of freedom
are dealt with in~\cite{fed10,fed13}. In these papers, some ``knife edges''
are added to the standard 2D hydrodynamical problem and the equations thus
obtained describe the hydrodynamical Chaplygin sleigh. However, generally speaking,
such an attachment of knife edges in an ideal fluid cannot be done from a mechanical
point of view, and the problem is described not by nonholonomic, but vakonomic
equations, see the review~\cite{bmb17}. In our problem, we use standard
equations of rolling of a rigid body in a magnetic field, and therefore
no additional nonrealistic hypotheses are required.

The paper is organized as follows. The rest of Section~\ref{bor-sec1}, devoted to Dirac's
development of Hamiltonian mechanics,
includes a discussion of the energy paradigm and deformation of the Poisson brackets.
In Section~\ref{bor-sec2} we construct a conformally Hamiltonian vector field describing the
motion of the Chaplygin sphere in an arbitrary solenoidal field. In the particular
case of magnetic  forces we obtain a nonholonomic analog of the Grioli problem~\cite{grioli47,grioli57}, see also~\cite{bb74} and the textbooks~\cite{bog91,bm01}.
Section 3 is devoted to discussion of the nonholonomic analog of the classical
Grioli and Barett\,--\,London problems for the Chaplygin sphere. For instance, we
consider the existence of an invariant measure and of integrals of motion for
a body which consists of dielectric and ferromagnetic parts. As a  by-product
we also obtain two new integrable cases of Hamiltonian rigid body dynamics in a
magnetic field.

\subsection{Magnetic field in a Dirac approach}

The passage from Lagrangian to Hamiltonian mechanics via a Legendre transformation is inadequate in several specific situations:
\begin{itemize}\itemsep=-1pt
  \item when the Lagrangian is a linear function in the velocity;
  \item when there are gauge or other unphysical degrees of freedom;
  \item when there are constraints that one wishes to impose in phase space.
\end{itemize}
For instance,  let us take  a particle with charge $q$  confined to the $xy$ plane
with a strong constant, homogeneous perpendicular magnetic field pointing in the $z$-direction with strength $B$~\cite{dun91}. In the limit of a very large magnetic field one may  drop the kinetic term to produce a simple approximate Lagrangian
\[
L=\dfrac{qB}{2c}(x\dot{y}-y\dot{x})-V(x,y).
\]
The corresponding equations of motion are
\[
\dot{y}=\dfrac{c}{qB}\dfrac{\partial V}{\partial x},\qquad \dot{x}=-\dfrac{c}{qB}\dfrac{\partial V}{\partial y},
\]
where $c$ is the speed of light in vacuum and $V(x,y)$ is an arbitrary external scalar potential. Note that this approximate Lagrangian is linear in the velocities, which is one of the conditions under which the standard Hamiltonian procedure breaks down.

Following the Hamiltonian procedure, however, the canonical momenta associated with the coordinates are now
\[
p_x = \frac{\partial L}{\partial \dot{x}} = -\frac{q B}{2c}y,\qquad
p_y = \frac{\partial L}{\partial \dot{y}} = \frac{q B}{2c}x,
\]
which are unusual in that they are not invertible to the velocities. Instead, they are constrained to be functions of the coordinates.
A Legendre transformation then produces the Hamiltonian
\[
 H(x,y, p_x, p_y) = \dot{x}p_x + \dot{y} p_y - L = V(x, y).
 \]
 It is easy to see that this Hamiltonian has no dependence on the momenta, which means that Hamilton's equations of motion  are
inconsistent. For a complete discussion, see~\cite{dun91}.

In~\cite{dir1,dir2,dir} Dirac argues that we should generalize the Hamiltonian somewhat analogously to the method of Lagrange multipliers and deform Poisson bracket in order to treat classical systems with
second-class constraints in Hamiltonian mechanics, and to thus allow them to undergo canonical quantization. It is an important part of Dirac's development of Hamiltonian mechanics to elegantly handle more general Lagrangians,
see also~\cite{flato} for a discussion of possible generalizations.

Let us briefly recall the main ideas of  Dirac's development of Hamiltonian mechanics.
In Newtonian mechanics the addition of the external force $F'$ changes the equation
of motion
\[
m\ddot{q}_i=F_i\to m\ddot{q}_i=F_i+F',\qquad i = 1,\ldots,n,
\]
where $q_i$ are coordinates on a configurational space.

In the Lagrangian formulation, we have the function $L(q_i, \dot{q}_i, t)$, where $q_i$
are  generalized coordinates and the equations of motion are
\[
\dfrac{\mathrm d}{\mathrm d t}\left(\dfrac{\partial L}{\partial \dot{q}_i}\right)-\dfrac{\partial L}{\partial q_i}=0\,.
\]
In order to change these equations, we have to shift $L$ by generalized potential $U(q_i,\dot{q}_i)$
\[
L(q_i,\dot{q}_i,t)\to L'(q_i,\dot{q}_i,t)=L(q_i,\dot{q}_i,t)-U(q_i,\dot{q}_i)\,.
\]
The Lagrangian $L(q_i, \dot{q}_i, t)$ is a function of the coordinates $q_i$, their time derivatives $\dot{q}_i$
and  time.

The basic idea of Hamilton's approach is to introduce generalized momenta
\[
p_i=\dfrac{\partial L}{\partial \dot{q}_i}.
\]
and Hamiltonian $H$ which is the Legendre transform of the
Lagrangian $L$ with respect to the $\dot{q}_i$ variables
\[
H(q_i,p_i,t)=\sum_{i=1}^n p_i\dot{q}_i -L(q_i, \dot{q}_i, t).
\]
Here $\dot{q}_i$ is eliminated from the right-hand side in favor of $p_i$. Thus, in Hamiltonian formalism
equations of motion are defined by the Hamiltonian function $H$ and by the
Poisson bivector $P$
\begin{equation}\label{ham-eq}
\dfrac{\mathrm d x_i}{\mathrm dt}=X_i,\qquad X=P\mathrm dH,
\end{equation}
where $x=(q_1,\ldots,q_n,p_1,\ldots,p_n)$ is a point in $2n$-dimensional phase space.
Together with  the Poisson bivector $P$ we  use the Poisson brackets defined by
\[
\{f(x),g(x)\}=(\mathrm df, P\mathrm d g),\qquad \mathrm df=\left(\frac{\mathrm df}{\mathrm dx_1},\frac{\mathrm df}{\mathrm dx_2},\ldots,\frac{\mathrm df}{\mathrm d x_n}\right),
\]
where $f$ and $g$ are two functions on the phase space and $(.,.)$ means a scalar product of two $2n$-dimensional vectors.

Addition of external force replaces the original momenta $p_i$ to generalized momenta
\[
p_i\to \tilde{p}_i=\dfrac{\partial L}{\partial \dot{q}_i}-\dfrac{\partial U}{\partial \dot{q}_i}=p_i-\dfrac{\partial U}{\partial \dot{q}_i}.
\]
and changes the Hamiltonian
\begin{equation}\label{ob-ham}
H(q_i,p_i,t)\to H'(q_i,p_i,t)=H(q_i,p_i,t)+U(q_i,p_i).
\end{equation}
If the original Hamiltonian $H$ is the total mechanical energy and the work done
by external forces is equal to zero, the total mechanical energy $H$ is altered and
the new Hamiltonian $H'$ has no mechanical meaning.

According to Dirac, we can preserve the energy paradigm by saying that the total
mechanical energy contains all the dynamical information and mathematical
Hamiltonian formalism by using deformation of the Poisson brackets.
In Dirac's approach the addition of nonconservative external force changes
the original equations of motion~$(\ref{ham-eq})$ to equations of the form
\[\dfrac{\mathrm d x_i}{\mathrm dt}=X'_i\,,\qquad X'=P'\mathrm dH\]
instead of equivalent equations
\[\dfrac{\mathrm d x_i}{\mathrm dt}=X'_i\,,\qquad X'=P\mathrm dH'\,,\]
which follow from the Legendre transformation of the generalized
Lagrangian $L'=L-U$. Having the equations of motion, however, is not the endpoint
for theoretical considerations. If one wants to canonically quantize a general
system or study its other deformations, then one needs the Dirac brackets.

Thus, in Hamiltonian formalism we can change equations of motion~$(\ref{ham-eq})$ by using
\begin{itemize}\itemsep=-1pt
  \item deformation of Hamiltonian $H\to H'$,
  \item deformation of the Poisson bivector $P\to P'$,
\end{itemize}
which both are related to transformation of momenta $p_i$ to generalized momenta
$\tilde{p}_i$. Deformations of the Poisson brackets associated with a magnetic
(solenoidal) field  are discussed in~\cite{mar07}.

In this paper we discuss the application of the Dirac idea in nonholonomial mechanics.
As an example, we consider a nonholonomic Chaplygin sphere in a magnetic (solenoidal)
field. In this case we have to study the composition of a known deformation
associated with reaction forces of the nonholonomic constraint~\cite{bm14} with
a known deformation associated with a magnetic field~\cite{casu11,mar07}.

\section{Deformations of the Poisson brackets}
\label{bor-sec2}

Let us consider a Lie\,--\,Poisson algebra $e^*(3)$ endowed with the linear Poisson
brackets
  \begin{equation}\label{e3}
  \bigl\{\gamma_i,\gamma_j\,\bigr\}=0.
 \quad
\bigl\{M_i,\gamma_j\,\bigr\}=\varepsilon_{ijk}\gamma_k ,
\quad
\bigl\{M_i,M_j\,\bigr\}=\varepsilon_{ijk}M_k,
\end{equation}
 where $\varepsilon_{ijk}$ is a complete antisymmetric tensor. The corresponding
Lie\,--\,Poisson bivector is equal to
 \begin{equation} \label{p-e3}
 P=\left(\begin{array}{cc}0& \Gamma\\  \Gamma &
 \mathrm M\end{array}\right).
 \end{equation}
Here $\gamma=(\gamma_1,\gamma_2,\gamma_3)$ and $M=(M_1,M_2,M_3)$ are vectors in three-dimensional
Euclidean space $\mathbb R^3$, which can be identified with $3\times3$ skew-symmetric
matrices in $so(3)$
\[
 \Gamma=\left(
                 \begin{array}{ccc}
                   0 & \gamma_3 & -\gamma_2 \\
                   -\gamma_3 & 0 & \gamma_1 \\
                   \gamma_2 & -\gamma_1 & 0
                 \end{array}
               \right),\quad
\mathrm M=\left(
                 \begin{array}{ccc}
                   0 & M_3 & -M_2 \\
                   -M_3 & 0 & M_1 \\
                   M_2 & -M_1 & 0
                 \end{array}
               \right),
\]
using the standard isomorphism of Lie algebras  $\left(\mathbb R^3,a\times b\right)$
and $\left(so(3),[A,B]\right)$, where $a\times b$ is a vector product and $[A,B]$ is a
matrix commutator.

Bivector $P$ has two Casimir functions:
\begin{equation}\label{caz0}
C_1=(\gamma,\gamma)\,,\qquad C_2=(\gamma,M),\quad P\mathrm dC_1=0\,,\quad P\mathrm dC_2=0,
\end{equation}
and, together with the Hamiltonian
\begin{equation}\label{ham}
 H=\dfrac12 (M,\omega)+V(\gamma),
\end{equation}
generates the Hamiltonian vector field
 \[X=P\mathrm dH.\]
If we identify coordinates  $M$ with the angular momentum vector of the rigid body,
$\gamma$ with the unit Poisson vector in  the rigid body frame and $H$ with the total
mechanical energy of the rigid body, then this vector field corresponds to the
standard Euler\,--\,Poisson equations describing the rotation of the rigid body
around a fixed point in a potential field
 \[
 \dot{\gamma}=\gamma\times\omega\,,\qquad \dot{M}=M\times \omega-\dfrac{\partial V}{\partial \gamma}\times \gamma.
 \]
Here $\omega=AM$ is the angular velocity vector,  $A=I^{-1}=\text{diag}(a_1,a_2,a_3)$
is an inverse matrix  to the tensor of inertia $I$. All the vectors are expressed
in the so-called body frame,  which is firmly attached to the rigid body,
its origin is at the center of mass of the body, and its axes coincide with
the principal inertia axes of the body~\cite{bog91,bm01}.

\subsection{Magnetic Poisson brackets}

Construction of the so-called magnetic Poisson structures which are deformations of the standard Lie\,--\,Poisson brackets on the various Lie algebras may be found in~\cite{mar07}.
Here we discuss only a partial deformation of the Poisson brackets~$(\ref{e3})$
on $e^*(3)$ which is related to the well-known transformation of
the original angular momentum
\begin{equation}\label{m-trans}
\varphi:\quad M_i\to\tilde{M}_i= M_i+c_i(\gamma_1,\gamma_2,\gamma_3),\qquad i=1,2,3,
\end{equation}
to the generalized angular momentum $\tilde{M}_i$, which allows us to present
the generalized Hamiltonian function
\[H'(q,p)=T(q,p)+V(q)+U(q,p)\] as a sum of generalized kinetic energy and potential
\[H'(q,p)=T'(q,p)+V(q),\]
see~\cite{bog91} for details and references therein.

Mapping~$(\ref{m-trans})$ reduces the original linear Lie\,--\,Poisson brackets~$(\ref{e3})$
to the following Poisson brackets:
 \begin{equation}\label{e3-b}
  \bigl\{\gamma_i,\gamma_j\,\bigr\}=0.
 \quad
\bigl\{M_i,\gamma_j\,\bigr\}=\varepsilon_{ijk}\gamma_k,
\quad
\bigl\{M_i,M_j\,\bigr\}=\varepsilon_{ijk}(M_k+b_k),
\end{equation}
where
\begin{equation}\label{eq-bc}
\begin{aligned}
b_1&= \left(\dfrac{\partial c_2}{\partial \gamma_2} + \dfrac{\partial c_3}{\partial \gamma_3}\right)\gamma_1 - \dfrac{\partial \left(\gamma_2 c_2+\gamma_3 c_3\right)}{\gamma_1} - c_1\,,
\\
b_2&=\left(\dfrac{\partial c_1}{\partial \gamma_1} + \dfrac{\partial c_3}{\partial \gamma_3}\right)\gamma_2
-\dfrac{\partial\left(\gamma_1c_1+\gamma_3c_3\right)}{\partial \gamma_2}- c_2\,,
\\
b_3&=\left(\dfrac{\partial c_1}{\partial \gamma_1} + \dfrac{\partial c_2}{\partial \gamma_2}\right)\gamma_3
 -\dfrac{\partial\left(\gamma_1c_1+\gamma_2c_2\right)}{\partial g_3}- c_3\,,
\end{aligned}
\end{equation}
so that
\begin{equation}\label{curl-b}
(\mbox{rot}\, b,\gamma)=(\nabla\times b,\gamma)=0.
\end{equation}
Here $b=(b_1,b_2,b_3)$ is a vector depending on coordinates $\gamma$.
Associated with the Poisson brac\-kets~(\ref{e3-b}) is the deformation of the
Poisson bivector, which has the form
 \begin{equation} \label{p-e3-b}
 P_\varphi=\left(\begin{array}{cc}0& \Gamma\\  \Gamma &
 \mathrm M+\mathrm B\end{array}\right),
 \qquad
\mathrm B=\left(
                 \begin{array}{ccc}
                   0 & b_3 & -b_2 \\
                   -b_3 & 0 & b_1 \\
                   b_2 & -b_1 & 0
                 \end{array}
               \right)\,.
\end{equation}
The corresponding deformations of the Casimir functions (\ref{caz0}) read as
\[
C_1=(\gamma,\gamma),\qquad C_2=(\gamma,M-c),\quad P_\varphi\mathrm dC_1=0\,,\quad P_\varphi\mathrm dC_2=0.
\]

Let us describe an inverse transformation $P_\phi\to P$ and suppose that we
have bivector $P_\varphi$~$(\ref{p-e3-b})$ with arbitrary functions $b_1,b_2$ and $b_3$.
This bivector satisfies the Jacobi condition only if these functions satisfy
condition~$(\ref{curl-b})$
\[
(\mbox{rot}\, b,\gamma)=(\nabla\times b,\gamma)=0.
\]
However, substituting functions $b_1,b_2$ and $b_3$ into~$(\ref{eq-bc})$, we obtain
an inconsistent system of differential equations with respect to functions $c_1,c_2$
and $c_3$ even if~$(\ref{curl-b})$ holds. These equations become compatible only
if we suppose that $P_\phi$ has two Casimir functions linear in momenta $M_i$.

\begin{lem}
Bivector $P_\phi$~$(\ref{p-e3-b})$ depending on arbitrary functions $b_1, b_2$ and $b_3$ of coordinates $\gamma_1,\gamma_2,\gamma_3$  is a Poisson bivector with two Casimir functions linear in $M_i$
 \begin{equation}\label{caz2-gen}
C_1=(\gamma,\gamma),\qquad C_2=(\gamma,M)+e(\gamma_1,\gamma_2,\gamma_3)\,
\end{equation}
if $b_1(\gamma_1,\gamma_2,\gamma_3)$ and $e(\gamma_1,\gamma_2,\gamma_3)$ are arbitrary functions, whereas
\begin{equation}\label{b-cond}
b_2=\dfrac{\gamma_2}{\gamma_1}\left( b_1(\gamma)  - \dfrac{\partial e(\gamma)}{\partial \gamma_1 )}\right)+ \dfrac{\partial e(\gamma)}{\partial \gamma_2},\qquad
b_3= \dfrac{\gamma_3}{\gamma_1}\left( b_1(\gamma) - \dfrac{\partial e(\gamma)}{\partial \gamma_1}\right)+ \dfrac{\partial e(\gamma)}{\partial \gamma_3},
 \end{equation}
 up to permutation of the indices of $b_k$.
\end{lem}

Now, substituting functions  $b_{1,2}$~$(\ref{b-cond})$ into Eqs.~$(\ref{eq-bc})$,
we obtain a compatible system of differential equations of second order on
$c_1, c_2$ and $c_3$. The generic solution of these equations may be obtained by
using a computer algebra system. For brevity we do not present this solution here.

The new Poisson bivector  $P_\phi$~$(\ref{p-e3-b})$ and the original mechanical energy
$H$~$(\ref{ham})$
generate the Hamiltonian vector field
 \[
 X_\varphi=P_\varphi\mathrm d H\,,\qquad  H=\dfrac12 (M,\omega)+V(\gamma),
\]
associated with the Euler\,--\,Poisson equations
   \[
 \dot{\gamma}=\gamma\times\omega,\qquad \dot{M}=(M+b)\times \omega-\dfrac{\partial V}{\partial \gamma}\times \gamma,
 \]
describing the rotation of the rigid body about fixed points in
a potential field with potential $V(\gamma)$ and in a solenoidal field $\gamma\times b$,
because
\[ \mbox{div}\,\gamma\times b=0\]
according to~$(\ref{curl-b})$.

At $b=B\gamma+\alpha$ the vector field $X_\varphi$ describes the dynamics of the
charged rigid body with stationary charge distribution (dielectric) rotating
about the fixed point in a constant magnetic field, for details about
this Grioli problem, see the textbooks~\cite{bog91,bm01}. In this case
the symmetric matrix $B$ describes an electric charge distribution, whereas the
vector $\alpha$ is a vector of gyrostatic momentum. Similar equations appear in the
description of an underwater vehicle when matrix $B$ describes a buoyancy distribution~\cite{mar07}.

Substituting into the definition of the vector field $X_\varphi$ various
Hamiltonians $H$ associated with the well-known integrable cases of
rigid body motion, one gets Euler, Lagrange, Kowalewski tops or
Clebsch  and Steklov\,--\,Lyapunov systems in a solenoidal field, but in the general case
the addition of a magnetic field destroys integrability of the original
dynamical systems  \cite{casu11,bm01,mar07,koz85}.

\subsection{Nonholonomic magnetic Poisson brackets}

In~\cite{bbm19,bts12,bol15,bm14,ts12,ts19} we construct linear in $M_i$ transformations of angular momentum
\begin{equation}\label{n-trans}
\psi:\qquad M_i\to\tilde{M}_i=f_i(\gamma)M_i+g_i(\gamma)
\end{equation}
which reduce the original Lie\,--\,Poisson bivector $P$ (\ref{p-e3})  to a Poisson bivector
$P_\psi$, which allows us to decompose the non-Hamiltonian vector field by
Hamiltonian vector fields. It allows us to use methods of Hamiltonian mechanics to
study non-Hamiltonian systems such as the Chaplygin sphere on a plane, a sphere
and a turntable, other generalizations of the Chaplygin sphere model,
the Routh sphere model, the Veselova system and many other nonholonomic
systems.

It is natural to study the decomposition of the mappings $\varphi$~$(\ref{m-trans})$
and $\psi$~$(\ref{n-trans})$ which corresponds to a simultaneous addition of the
external nonconservative forces  and reaction forces associated with nonintegrable
constraints. In this note we consider the transformation of $M_i$ variables
\begin{equation}\label{m-trans2}
\psi:\quad\left\{
\begin{aligned}
M_1\to \tilde{M}_1&= g\left(M_1-\dfrac{\beta \gamma_1}{\gamma_1^2+\gamma_2^2}\right)
+\dfrac{\alpha \gamma_1}{(\gamma,\gamma)}\left(1+\dfrac{\gamma_3^2}{\nu}\right),
\\
M_2\to \tilde{M}_2&=g\left(M_2-\dfrac{\beta \gamma_2}{\gamma_1^2+\gamma_2^2}
\right)+\dfrac{\alpha \gamma_2}{(\gamma,\gamma)}\left(1+\dfrac{\gamma_3^2}{\nu}\right),
\\
M_3\to \tilde{M}_3&=gM_3+
\dfrac{\alpha \gamma_3}{(\gamma,\gamma)}\left(1-\dfrac{\gamma_1^2+\gamma_2^2}{\nu}\right),
\end{aligned}
\right.
 \end{equation}
 where
  \[ \alpha =(\gamma,M),\quad  \beta = (\gamma,L) \quad \mbox{and }\quad  \nu=\gamma_1^2+\gamma_2^2-d(\gamma,\gamma)(a_1\gamma_1^2+a_2\gamma_2^2)\,.\]
This mapping and the corresponding deformation of the Lie\,--\,Poisson bivector $P$~$(\ref{p-e3})$ depend on parameter  $d$ and the diagonal matrix $A=\text{diag}(a_1,a_2,a_3)$,
which determine the function
\[
g=\sqrt{1-d ( \gamma, A \gamma)}.
\]
Mapping $\psi$~$(\ref{m-trans2})$ reduces the original Lie\,--\,Poisson bivector $P$~$(\ref{p-e3})$ to the following Poisson bivector:
\begin{equation}\label{ch-p}
 P_\psi=g\,P-\frac{d}{g}\,(M,  A\gamma)\left(\begin{array}{cc}0&0\\ 0&
  \Gamma\end{array}\right),
\end{equation}
and preserves the form of the original Casimir functions~$(\ref{caz0})$
\[
C_1=(\gamma,\gamma),\quad C_2=(\gamma,M),\quad P_\psi\mathrm dC_1=0,\quad P_\psi\mathrm dC_2=0.
\]
Let us identify $M=(M_1,M_2,M_3)$ with the angular momentum vector of the  Chaplygin
sphere with respect to a contact point and the vector $\omega=(\omega_1,\omega_2,\omega_3)$
\begin{equation}\label{w-ch}
\omega=A_\gamma\,M,\qquad A_\gamma=A+\dfrac{dA\, \gamma\otimes \gamma\, A}{g^2},\qquad a_k=(I_k+d)^{-1}
\end{equation}
with the angular velocity vector of the rolling sphere. Its mass, inertia
tensor and radius will be denoted by $m$, $I = \mathrm{diag}(I_1, I_2, I_3 )$ and $b$, respectively.
Parameter $d$ involves the mass and radius of the Chaplygin sphere $d=mb^2$,
see~\cite{ts11,bm14} for details.

In this case the bivector $P_\psi$, together with unvaried total mechanical energy  $H$~$(\ref{ham})$
\[H=\dfrac{1}{2}(M,\omega)+V(\gamma)\]
where $\omega$ is given by~$(\ref{w-ch})$, generates the conformally Hamiltonian
vector field
\[
X_\psi=g^{-1}P_\psi\mathrm d H,\qquad  H=\dfrac12 (M,\omega)+V(\gamma),
\]
endowed with the invariant measure
 \begin{equation}\label{mu}
 \mu=g\,\mathrm d\gamma\,\mathrm dM
 \end{equation}
and having three first integrals $C_{1,2}$ and $H$. The corresponding equations
of motion
\[
  \dot \gamma= \gamma\times \omega,\qquad \dot M=M\times \omega
   - \dfrac{\partial V}{\partial \gamma}\times \gamma\,
\]
describe the rolling of the Chaplygin sphere on a plane in a potential field~\cite{ts11,bm14}.

By adding an external nonconservative force we have to replace these
equations
with the equations
 \begin{equation} \label{ch-eqm}
 \dot{\gamma}=\gamma\times\omega,\qquad \dot{M}=M\times \omega+b\times \omega-\dfrac{\partial V}{\partial \gamma}\times \gamma,
 \end{equation}
where the angular velocity vector is given by~$(\ref{w-ch})$.

\begin{thr}
 Equations~$(\ref{ch-eqm})$ have an invariant measure
 \[ \mu=g^{-1}\,\mathrm d\gamma\,\mathrm dM.\]
It means that imposing the solenoidal field on the nonholonomic Chaplygin sphere
preserves both the total mechanical energy~$(\ref{ham})$ and the invariant measure~$(\ref{mu})$.
\end{thr}
The proof is a trivial observation that the equation on the last Jacoby multiplier
\[\mathrm{div} \rho(\gamma_1,\gamma_2,\gamma_3) X_{\psi\varphi}=0\]
is independent of the functions $b_k$ from the definition of the bivector $P_\varphi$. So,
the solution of this equation can be used in a standard construction of the invariant measure simultaneously  for the vector fields $X_\psi$ and $X_{\psi\varphi}$.

Composition $\psi\circ\varphi$ of the mappings  $\varphi$~$(\ref{m-trans})$ and
$\psi$~$(\ref{m-trans2})$ reduces the original Lie\,--\,Poisson bivector $P$~$(\ref{p-e3})$ to
the following Poisson bivector:
\begin{equation}\label{ch-b-p}
P_{\psi\varphi}=g\,P_\varphi-\frac{d}{g}\,(M,  A\gamma)\left(\begin{array}{cc}0&0\\ 0&
  \Gamma\end{array}\right).
\end{equation}
In fact, in order to get this expression, we have to replace $P$ with $P_\varphi$ in $P_\psi$~$(\ref{ch-p})$ and do not change $M_i$ in the second term.
As above, this bivector satisfies the Jacobi condition only if
condition~$(\ref{curl-b})$ holds and has the same Casimir function as the
bivector $P_\varphi$~$(\ref{p-e3-b})$
\[
C_1=(\gamma,\gamma),\qquad C_2=(\gamma,M+c),\quad P_{\psi\varphi}\mathrm dC_1=0,\quad P_{\psi\varphi}\mathrm dC_2=0.
\]

\begin{thr}
The equations of motion~$(\ref{ch-eqm})$ in which the angular velocity vector $\omega$
is given by~$(\ref{w-ch})$ are conformally Hamiltonian equations associated with
the following vector field:
\begin{equation}\label{ch-x-b}
X_{\psi\varphi}=g^{-1}P_{\psi\varphi}\mathrm d H,\qquad H=\dfrac12 (M,\omega)+V(\gamma),
\end{equation}
generated by the total mechanics energy $H$ and by deformation of the Lie\,--\,Poisson
brackets~$(\ref{e3})$.
\end{thr}

The proof consists of a direct verification.

Summing up, we prove the efficiency and relevance of the Dirac method in
non-Hamiltonian mechanics  using the nonholonomic  Chaplygin sphere as an example.
In a similar manner we can use compositions of the various known deformations of
the Poisson brackets to construct equations of motion for other nonholonomic
systems~\cite{bbm19,bts12,bol15,bm14,ts11,ts12,ts19}.

\section{Chaplygin sphere and Barnett-London effect}
\label{bor-sec3}

It is well known that a magnetic body can be magnetized when rotating
about one of its axes~\cite{lan}. This magnetization by rotation was predicted
and confirmed for ferromagnetic bodies by Barnett in 1915~\cite{barn}.
The same phenomenon  was predicted in 1933 by Becker, Heller and Sauter~\cite{beck}
for a perfect conductor set into rotation.  Subsequently London predicted that the
same final state should result when a rotating normal metal is cooled into the
superconducting state~\cite{lond}. The resulting magnetic moment will depend on the
shape of the body and is called the ``London moment''. A~modern discussion of
the dynamical understanding of the Meissner effect and the London moment may be
found in~\cite{hir19}, where the reader can also find references on
experimental verifications of this effect for a variety of superconductors.

The 1980s witnessed many publications devoted to the problem of rigid body motion
taking the Barnett\,--\,London
effect into account~\cite{kob81,koz85,sam84,vit, yeg83,yeh85}, see also the recent
review~\cite{hus17} and references therein.

In this section we consider the motion of the nonholonomic Chaplygin sphere acted upon
by a magnetic field, taking the Barnett\,--\,London effect into account.
The corresponding equations of motion are equal to
 \begin{equation}\label{eq-g}
\dot{M}=(M+B\gamma+\alpha)\times \omega+\left(C\,\omega-\dfrac{\partial V}{\partial \gamma}\right)\times \gamma,\qquad \dot{\gamma}=\gamma\times\omega.
 \end{equation}
Here the entries of the angular velocity vector $\omega=(\omega_1,\omega_2,\omega_3)$
are arbitrary functions of coordinates  $\gamma$ and $M$,  $B$ and $C$ are $3\times 3$
symmetric matrices,  $\alpha=(\alpha_1,\alpha_2,\alpha_3)$ is
the constant vector of gyrostatic moment, and the potential $V(\gamma)$ is a function
on coordinates $\gamma$.

If $C=0$ and
\begin{equation}\label{w}
\omega= AM\,,\qquad A=\left(
                 \begin{array}{ccc}
                   a_1 & 0 & 0 \\
                   0 & a_2 & 0 \\
                   0 & 0 & a_3
                 \end{array}
               \right),
\end{equation}
the equations of motion are Hamiltonian equations~\cite{bog91}. In the case of the
Grioli problem these equations describe
the rotation of a dielectric rigid body with gyrostatic moment $\alpha\neq 0$
about a fixed point in a permanent homogeneous magnetic field and in
a potential field~\cite{grioli47,grioli57}. In this case the matrix  $B$ determines
charge distribution, see~\cite{bb74,bog91,bm01} for details.
For similar equations appearing in other mechanical models, see the textbooks~\cite{mr94,mar07}.

 If $\alpha=0$, $B=0$ and
\[
\omega= AM,
\]
Eqs.~$(\ref{eq-g})$ are non-Hamiltonian equations describing the rotation of a
ferromagnetic rigid body about a fixed point in a permanent homogeneous magnetic
field and in a potential field, taking into account magnetization by rotation,
see~\cite{sam84,koz85} and Appendix~D in~\cite{bm01}. In this case, the matrix~$C$
diagonal in the principal system
of coordinates of the rigid body determines the anisotropy of ferromagnetic.
Below we will consider an arbitrary symmetric matrix $C$.

If $\alpha=0$, $B=C=0$ and the angular velocity vector is given by~$(\ref{w-ch})$
\[
\omega=A_\gamma\,M,\qquad A_\gamma=A+\dfrac{dA\, \gamma\otimes \gamma\, A}{g^2},\qquad g=\sqrt{1-d (\gamma, A \gamma)},
\]
the equations of motion~$(\ref{eq-g})$ are conformally Hamiltonian equations
describing the motion of the nonholonomic Chaplygin sphere on a plane,
which we have considered in the previous section.

It is quite natural to study the motion of the Chaplygin sphere in a magnetic field
when this sphere is made of a dielectric $B\neq 0$ or ferromagnetic material
$C\neq 0$.
So, below  we  substitute $\omega=A_\gamma M$ into~$(\ref{eq-g})$ and find out which
properties of the original nonholonomic system are preserved by imposing a magnetic
field. This substitution  $\omega=A_\gamma M$ can also be interpreted as imposing a
suitable nonholonomic constraint  on the well-studied dielectric and ferromagnetic
rigid body in a magnetic field.

\subsection{Invariant measure and quadratic first integrals}

The equations of motion~$(\ref{eq-g})$ have two so-called geometric first integrals
which are independent of the form of the angular velocity vector
 \begin{equation}\label{int-c12}
J_1=(\gamma,\gamma)\,,\qquad J_2=(\gamma,M+\alpha)+\dfrac12(\gamma,B\gamma).
\end{equation}
We can identify these first integrals with the Casimir functions of some deformations
of the Lie\,--\,Poisson bivector~$(\ref{e3-b})$, but in the generic case the corresponding
vector field
\[
\dfrac{dx_i}{dt}=X_i,\quad i=1 \ldots 6, \quad x=(\gamma,M),
\]
is not a Hamiltonian vector field.

\begin{thr}
\label{teo3}
At $\omega=AM$ the vector field $X$~$(\ref{eq-g})$ has an invariant measure
\[\mu=\mathrm d\gamma \mathrm d M\]
only if the matrix $C$ is a diagonal matrix
\begin{equation}\label{c-cond1}
C=\left(
                 \begin{array}{ccc}
                   c_1 & 0 & 0 \\
                   0 & c_2 & 0 \\
                   0 & 0 & c_3
                 \end{array}
               \right).
\end{equation}
At $\omega=A_gM$ the vector field $X$~$(\ref{eq-g})$ has an invariant measure
 \[
\mu=g^{-1}\mathrm d\gamma \mathrm d M,\qquad g=\sqrt{1-d ( \gamma, A \gamma)},
\]
only if the matrix $C$ is a diagonal matrix with entries satisfying the Clebsch
type condition
\begin{equation}\label{c-cond2}
\dfrac{c_2 - c_3}{a_1} +\dfrac{c_3-c_1}{a_2} +\dfrac{c_1 - c_2}{a_3}=0.
\end{equation}
\end{thr}
So, at $\omega=AM$ the well-known physical condition of diagonalization of $C$
in the principal system
of coordinates of the rigid body coincides with the condition of the existence of
an invariant measure.

Now let us come back to the generic symmetric matrix $C$ and substitute the
total mechanical energy
\[
H=\dfrac12(\omega,M)+V(\gamma)
\]
into the equation $X(H)=0$. Solving the resulting equation with respect to the
entries of matrices $B,C$ and the entries of vector $\alpha$, one gets the following
theorem.

\begin{thr}
The total mechanical energy is a first integral of Eqs.~$(\ref{eq-g})$ at
$\omega=AM$ and at $\omega=A_gM$ only if the matrix $B$ and the vector $\alpha$ are
arbitrary
   \[
   B=\left(
       \begin{array}{ccc}
         b_{11} & b_{12} & b_{13} \\
         b_{12} & b_{22} & b_{23} \\
         b_{1,3} & b_{23} & b_{33} \\
       \end{array}
     \right),\qquad \alpha=(\alpha_1,\alpha_2,\alpha_3),
\]
and the matrix $C$ is proportional to the unit matrix
\[C=\lambda\,E,\quad E=\left(
    \begin{array}{ccc}
      1& 0 & 0 \\
      0 & 1 &0 \\
      0 & 0 & 1 \\
    \end{array}
  \right),\qquad \lambda \in \mathbb R.
  \]
\end{thr}
So, if the total mechanical energy is a first integral of Eqs.~$(\ref{eq-g})$,
then these equations have an invariant measure according to Theorem~\ref{teo3}.
It is quite obvious because the total energy is altered only if the work done by
external forces is equal to zero.

Substituting these conditions into the second part of the equations of motion~$(\ref{eq-g})$, we can remove the ``non-Hamiltonian''\, term $C\omega\times \gamma$
from these equations:
\begin{equation}\label{trans-koz}
\begin{aligned}
\dot{M}&=\Bigl(M+B\gamma+\alpha\Bigr)\times \omega+\left(C\,\omega-\dfrac{\partial V}{\partial \gamma}\right)\times \gamma
=(M+B\gamma+\alpha)\times \omega+\left(\lambda \omega-\dfrac{\partial V}{\partial \gamma}\right)\times \gamma \\
&=\Bigr(M+(B-\lambda E)\gamma+\alpha\Bigl)\times \omega-\dfrac{\partial V}{\partial \gamma}\times \gamma,
\end{aligned}
\end{equation}
see~\cite{koz85}. The corresponding vector field $X$ is a Hamiltonian vector
field at $\omega=AM$ and a conformally Hamiltonian vector field in the nonholonomic
case $\omega=A_\gamma M$, see the previous section.

Now let us come back to the generic symmetric matrix $C$ and substitute the
square of the angular momentum
\[
M^2=(M,M)=M_1^2+M_2^2+M_3^2\,
\]
into the equation $X(M^2)=0$. Solving the resulting equation with respect to the entries of
the matrices $B,C$ and the entries of the vector $\alpha$, one gets the following
theorem.

\begin{thr}
The square of the angular momentum is a first integral of Eqs.~$(\ref{eq-g})$ only
if  $\alpha=0$ and $V(\gamma_1,\gamma_2,\gamma_3)=0$ .

At $\omega=AM$ the matrix $C$ has to be a diagonal matrix~$(\ref{c-cond1})$, whereas
the matrix $B$ is equal to
\[
 B=\left(
    \begin{matrix}
      \dfrac{a_2c_2 - a_3c_3}{a_2 - a_3}& 0 & 0 \\
      0 & \dfrac{a_3c_3 - a_1c_1}{a_3 - a_1} &0 \\
      0 & 0 & \dfrac{ a_1c_1 - a_2c_2}{a_1 - a_2}\\
    \end{matrix}
  \right).
 \]

At $\omega=A_\gamma M$ the matrix $C$ has to be a diagonal matrix with
entries satisfying~$(\ref{c-cond2})$, whereas the matrix $B$ is proportional to
the unit matrix, for instance,
\[
 B=\dfrac{a_2c_2 - a_3c_3}{a_2 - a_3}\left(
    \begin{array}{ccc}
      1& 0 & 0 \\
      0 & 1 &0 \\
      0 & 0 & 1\\
    \end{array}
  \right),
 \]
up to permutation of indices.
\end{thr}
So, if the square of the angular momentum is a first integral of Eqs.~$(\ref{eq-g})$,
then these equations have an invariant measure according to Theorem~\ref{teo3}.

\subsection{Old and new partial cases of integrability by quadratures}

Let us consider a dynamically symmetric rigid body at $a_1=a_2$. In the
nonholonomic case,
i.\,e., at $\omega=A_\gamma M$, the first integral  linear in $M_i$ does not exist.

  At $\omega=AM$  Eqs.~$(\ref{eq-g})$ have a standard invariant measure
 \[\mu=\mathrm d\gamma \mathrm d M\]
only if
\begin{equation}\label{c-cond3}
C=\left(
                 \begin{array}{ccc}
                   c_{11} & c_{12} & 0 \\
                   c_{12} & c_{22} & 0 \\
                   0 & 0 & c_{33}
                 \end{array}
               \right),
\end{equation}
in contrast with Theorem~3 for the generic case.\goodbreak

At $V=f(\gamma_3)$ Eqs.~$(\ref{eq-g})$ have also a first integral which is
a function linear in $M_3$
\[
K=M_3 - c_{1 3}\gamma_1-c_{23}\gamma_2 + (c_{11}-b_{2} )\gamma_3
\]
if $\alpha=(0,0,\alpha_3)$, the matrix $C$ is an arbitrary symmetric matrix
  \[C=\left(
       \begin{array}{ccc}
         c_{11} & c_{12} & c_{13} \\
         c_{12} & c_{22} & c_{23} \\
         c_{13} & c_{23} & c_{33} \\
       \end{array}
     \right),
\]
and the matrix $B$ has the following form:
\[
B=\left( \begin{array}{ccc}
       b_1+c_{22} & 0 & 0 \\
       0 & b_1+c_{22} & 0 \\
       0 & 0 & b_3+c_{33} \\
     \end{array}
   \right) - C.
   \]
In contrast with the first integrals quadratic in momenta $M_i$
from Theorems 4 and 5 the existence of a linear integral $K$
   is not related to the existence of an invariant measure. For
instance, if $c_{13}\neq 0 $, there is a first integral $K$ linear in momenta,
but there exists no invariant measure.

Equations of motion~$(\ref{eq-g})$ have two so-called geometric first integrals~$(\ref{int-c12})$, which are independent of the form of the angular velocity vector.
In order to get a system completely integrable by quadratures, we have to find
an invariant measure and two independent first integrals according to the
Euler\,--\,Jacobi theorem. We can easily do it by combining conditions for the
existence of an invariant measure, mechanical energy $H$, square of angular  momenta
$M^2$ and a first integral $K$.

For instance,  condition~$(\ref{c-cond1})$ for the invariant measure leads to the
standard form of the first integral
\[
K=M_3 + (c_{11}-b_{1})\gamma_3.
\]
By adding the condition from Theorem 4 one gets the integrable case from~\cite{sam84}  with integrals of motion $K$ and
\[
H=\dfrac{1}{2}\left( a_1M_1^2 +a_1M_2^2 +a_3 M_3^2\right) + V(\gamma_3).
\]
Since $C=\lambda E$,
transformation~$(\ref{trans-koz})$ reduces this system to the well-known integrable
Kirchhoff case  in the Grioli problem
at
 \[B=\left(
    \begin{array}{ccc}
     b_1 & 0 & 0 \\
      0 &b_1  & 0 \\
      0 & 0 & a_3 \\
    \end{array}
  \right)\]
with gyrostatic moment $\alpha=\mu(0,0,1)$~\cite{bog91,kir,koz85}.

By adding the condition from Theorem 5 one gets an integrable case
with integrals of motion $K$ and
\[
M^2=M_1^2 +M_2^2 + M_3^2
\]
at
\[
C=\left(
    \begin{array}{ccc}
     a_3c_3+\frac{b_1(a_1 - a_3)}{a_1}  & 0 & 0 \\
      0 &a_3c_3+\frac{b_1(a_1 - a_3)}{a_1}  & 0 \\
      0 & 0 & a_1c_3 \\
    \end{array}
  \right),\quad\mbox{so that}\quad B=\left(
    \begin{array}{ccc}
     b_1 & 0 & 0 \\
      0 &b_1  & 0 \\
      0 & 0 & a_3 \\
    \end{array}
  \right).
\]
A particular case of this integrable system is discussed in Appendix D~\cite{bm01}.
We suppose that there is a change of variables which reduces this  system to
the same  integrable Kirchhoff case of rigid body motion.

There are also two independent first integrals which are second-order
polynomials in momenta~$M_i$ if  we have a completely symmetric rigid body
\[  a_1=a_2=a_3\,.\]
In this particular case the vector field $X$~$(\ref{eq-g})$ always has an invariant
measure
 \[\mu=\mathrm d\gamma \mathrm d M\]
for $\omega=AM$ or $\omega=A_\gamma M$ without any restriction on the entries
of the symmetric matrix $C$.

In this particular case at $\omega=AM=aM$, $\alpha=0$ and  $V(\gamma)=0$
Eqs.~$(\ref{eq-g})$
\begin{equation}\label{koz-eqm}
a^{-1}\dot{M}=B\gamma \times M+C\,M\times \gamma,\qquad a^{-1}\dot{\gamma}=\gamma\times M
 \end{equation}
 involve two arbitrary symmetric matrices $B$ and $C$.  Let us diagonalize
the matrix $C$  using rotations of the vectors $M$ and $\gamma$
\[C=diag(c_1,c_2,c_3).\]
It is easy to see that Eqs.~$(\ref{eq-g})$ with the diagonal matrix $C$
have the following first integral:
\[
F_1=\lambda_1M_1^2+\lambda_2M_2^2+\lambda_3M_3^2=(M,\Lambda M),\qquad \Lambda=\left(
                                                                                  \begin{array}{ccc}
                                                                                    \lambda_1 & 0 & 0 \\
                                                                                    0 & \lambda_2 & 0 \\
                                                                                    0 & 0 & \lambda_3 \\
                                                                                  \end{array}
                                                                                \right)
\]
 if
 \begin{equation}\label{b-bl}
 B=\left(
     \begin{matrix}
       \dfrac{\lambda_2c_3 -\lambda_3c_2}{\lambda_2 - \lambda_3} & 0 & 0 \\
       0 & \dfrac{\lambda_1c_3 - \lambda_3c_1}{\lambda_1 - \lambda_3} & 0 \\
       0 & 0 & \dfrac{\lambda_1c_2 -\lambda_2c_1}{\lambda_1 - \lambda_2} \\
        \end{matrix}
   \right).
 \end{equation}
 Here $\lambda_k\in \mathbb R$ are arbitrary parameters.

 If we impose additional restrictions on the matrices $B$, $C$ and $\Lambda$, we
obtain two integrable cases of rigid body motion in a magnetic field. In the first case
the additional restriction has the following form:
\[
(\lambda_2 - \lambda_3)c_1+(\lambda_3-\lambda_1)c_2+(\lambda_1-\lambda_2)c_3=0\,.
\]
If we solve this condition with respect to $c_1$
\[c_1=\dfrac{(\lambda_1- \lambda_3)c_2 - (\lambda_1 - \lambda_2)c_3}{\lambda_2 - \lambda_3}\]
and substitute the resulting expression  into the matrix $B$~$(\ref{b-bl})$, we see
that the matrix $B$ is a unit matrix  up to a scalar factor
\[
B=\dfrac{\lambda_2c_3 - \lambda_3c_2}{\lambda_2 - \lambda_3}\left(
    \begin{array}{ccc}
      1& 0 & 0 \\
      0 & 1 &0 \\
      0 & 0 & 1\\
    \end{array}
  \right).
 \]
It allows us to put $B=0$,  i.\,e., $c_k=\lambda_k$,  in Eqs.~$(\ref{koz-eqm})$
without loss of generality. The corresponding  first integral was found by V.\,V. Kozlov~\cite{koz85}
  \[
  F_2=(M,M)-2(M,C\gamma)+\left(\frac{\mbox{tr}\,(C-B)}{\mbox{tr}\Lambda}\right)^{2}\mbox{det}\Lambda\,
   (\gamma,\Lambda^{-1}\gamma).
 \]
In~\cite{ves86} Veselova proved that this system is equivalent to the well-known integrable Clebsch case for the Kirchhoff equations after some unobvious change of variables.

In the second case the additional restriction looks like
\[
(\lambda_2 - \lambda_3)\lambda_1c_1+(\lambda_3-\lambda_1)\lambda_2c_2+(\lambda_1-\lambda_2)\lambda_3c_3=0.
\]
If we solve this condition with respect to $c_1$
\[
c_1=\dfrac{(\lambda_1-\lambda_3)\lambda_2c_2 - (\lambda_1 - \lambda_2)\lambda_3c_3}{\lambda_1(\lambda_2 -\lambda_3)}\,
\]
and substitute $c_1$  into the matrix $B$~$(\ref{b-bl})$, we obtain the
following matrix:
\[
B=\dfrac{\lambda_2c_3 - \lambda_3c_2}{\lambda_2 - \lambda_3}\left(
    \begin{array}{ccc}
      1& 0 & 0 \\
      0 & 1 &0 \\
      0 & 0 & 1\\
    \end{array}
  \right)+\dfrac{c_2 - c_3}{(\lambda_2 -\lambda_3)\lambda_1}
  \left(\begin{array}{ccc}
      0& 0 & 0 \\
      0 &  (\lambda_1 -\lambda_2)\lambda_3&0 \\
      0 & 0 &(\lambda_1- \lambda_3)\lambda_2 \\
    \end{array}
  \right),
\]
which cannot be removed from Eqs.~$(\ref{koz-eqm})$. The additional first integral
is equal to
\[
F_2=(M,M)-2(M,C\gamma)+\frac{\mbox{tr}\,(C-B)}{\mbox{tr}\Lambda^{-1}}\,
   (C \gamma,\Lambda^{-1}\gamma).
\]
We suppose that there is a counterpart of the Veselova transformation which
reduces this  system to the well-known integrable Steklov\,--\,Lyapunov  case
of rigid body motion. We plan to discuss this transformation in a forthcoming
publication.

\section{Conclusion}
\label{bor-sec4}

Thus, using the Chaplygin sphere as an example, we have proved the possibility and
advisability of using Dirac's ideas not only in Hamiltonian classical and quantum mechanics,
but also in nonholonomic mechanics. In addition, we have obtained a number of new
results for the motion of the Chaplygin sphere on a plane in a constant magnetic
field under the assumption that the sphere is made of a dielectric $B\neq 0$ and
ferromagnetic $C\neq 0$ material. In a similar manner, using deformations
of Poisson brackets and corresponding Poisson maps, one can also study
various generalizations of the motion of the Chaplygin sphere~\cite{bbm19,ts19} and
other nonholonomic systems~\cite{bts12,bol15,bm14,ts12,ts14} in a magnetic field.

As a development of the problems considered here, we also point out the possibility
of introducing inhomogeneous and nonstationary magnetic fields~\cite{hir19, kob81,urm98}.
The first possibility is closely related to the Levitron theory and will allow the sphere, for example,
to jump without additional mechanical actions. By contrast, the nonstationary
field can in principle allow the sphere to be accelerated to the necessary velocity
or the chaotic regime to be reached, which is
diagnosed from intricate or even diffused motion of the contact point. All these processes
should be studied from the viewpoint of mobile robots. An excellent review
of these can be found in the dissertation~\cite{y17}.

A proof of unbounded growth of energy
under periodic oscillations of the rotor inside the Chaplygin sleigh and
of bounded growth in the presence of dissipation is given in~\cite{biz19}.
An unbounded acceleration for the Chaplygin sphere is discussed in~\cite{bor19}.
We note that the acceleration mechanism of spheres is still not understood.
The magnetic field, which is a complex gyroscopy, will allow progress in solving this
problem.

In  a  number  of  instruments  utilizing  a  contact less suspension
system  a  rotor  spins  at  high  speed  in  a  suspension magnetic  field.
The  design  of  such  instruments  is  feasible only  if  the  properties  of
bodies  rotating  in  a  magnetic  field are  known~\cite{hir19,kob81,urm98}.
We could enrich this theory by taking into account various nonholonomic effects
modelling
friction.

The work of A.\,V.\,Tsiganov (Sections~\ref{bor-sec1},~\ref{bor-sec2}) is supported by the Russian Science Foundation (project no.~19-71-30012) and performed in Steklov Mathematical Institute of Russian Academy of Sciences.
The work of  A.\,V.\,Borisov (Sections~\ref{bor-sec3},~\ref{bor-sec4}) was carried out within the framework of the state assignment of the Ministry of Education and Science of Russia (1.2404.2017/4.6).

\end{document}